\documentclass[showpacs,showkeys,twocolumn,nofootinbib]{revtex4-1}

\usepackage{graphicx}
\usepackage{amsmath}
\usepackage{amssymb}
\usepackage{slashed}

\newcommand{\Tr}{\mathop{\rm Tr}\nolimits}
\newcommand{\e}{\ensuremath{\mathrm{e}}}
\newcommand{\I}{\ensuremath{\mathrm{i}}}
\renewcommand{\d}{\ensuremath{\mathrm{d}}}
\newcommand{\D}{\ensuremath{\mathrm{D}}}
\newcommand{\T}{\ensuremath{\mathrm{T}}}
\newcommand{\eL}{\mathcal{L}}
\newcommand{\C}{\ensuremath{c}}
\newcommand{\trip}{\ensuremath{\mathbf{3}}}
\newcommand{\atrip}{\ensuremath{\overline{\mathbf{3}}}}
\newcommand{\group}[1]{\mathrm{#1}}

\def\im{\I}
\def\beginm{\left(\begin{array}}
\def\endm{\end{array}\right)}

\begin{document}

\title{Masses by gauge flavor dynamics}

\author{Petr Bene\v s}
\author{Ji\v r\'{\i} Ho\v sek} \email{hosek@ujf.cas.cz}
\author{Adam Smetana}
\affiliation{Department of Theoretical Physics, Nuclear Physics Institute, Academy of Sciences of the Czech Republic, 25068 \v Re\v z (Prague), Czech Republic}

\begin{abstract}
We gauge the experimentally observed flavor (family) index of chiral lepton and quark fields and argue that the resulting non-vectorial $\group{SU}(3)_{\mathrm{F}}$ dynamics completely self-breaks. This breakdown generates fermion masses, which in turn trigger electroweak symmetry breaking (EWSB). Suggested asymptotically free dynamics with an assumed non-perturbative infrared fixed point has just one free parameter and is therefore either right or plainly wrong. Weak point of field theories strongly coupled in the infrared, unfortunately, is that there is no reliable way of computing their spectrum. Because of its rigidity the model provides, however, rather firm theoretically safe experimental predictions without knowing the spectrum: First, anomaly freedom fixes the neutrino sector which contains almost sterile neutrino states. Second, global symmetries of the model, spontaneously broken by fermion masses imply the existence of a fixed pattern of (pseudo-)axions and (pseudo-)majorons. It is gratifying that the predicted both sterile neutrinos and the pseudo-Nambu--Goldstone bosons are the viable candidates for dark matter.
\end{abstract}




\maketitle

\section{Introduction}

One of the most persistent issues in the particle physics has been in last decades the problem of electroweak symmetry breaking (EWSB) and fermion mass generation. Most of the current approaches can be divided into two main categories. The first category contains weakly coupled models, in which the electroweak symmetry is broken typically by condensates of some electroweakly charged scalar fields. The scalar condensation translates the bare scalar mass parameters, present already at the level of Lagrangian, into masses of gauge bosons and fermions \cite{Higgs:1964pj,*Englert:1964et,*Guralnik:1964eu}. Most notable members of this group are the Standard Model (SM) \cite{Glashow:1961tr,*Salam:1968rm,*Weinberg:1967tq}, the Two Higgs Doublet Model \cite{Lee:1973iz} and the Minimal Supersymmetric Standard Model \cite{Nilles:1983ge,Haber:1984rc,Chung:2003fi}.

The second category consists of models where the EWSB is achieved dynamically, typically due to fermion condensates in analogy with superconductors \cite{Bardeen:1957mv,Nambu:1961tp,Freundlich:1970kn}
. The condensation is usually driven by some new \emph{strong} and \emph{chiral} gauge dynamics. The mass scale of the condensate is not present at the level of Lagrangian but instead it is generated by dimensional transmutation of the running gauge coupling constant. In order to manifest the scale as particle masses the corresponding gauge dynamics must not be stable in the infrared \cite{Stern:1976jg}. The \emph{asymptotically free} theories are suitable candidates. The leading representatives of this category are various Extended Technicolor (ETC) models \cite{Dimopoulos:1979es,Eichten:1979ah,Farhi:1980xs,[{For a review, see }][{}]Accomando:2006ga}. They all introduce new fermions, charged under both the electroweak and the new strong gauge dynamics.

Recent suggestion \cite{Hosek:2009ys,Hosek:NagoyaProceeding} of one of us belongs to the second category. Although it resembles in some aspects the ETC models, it is still substantially different. The basic idea, though not completely new \cite{Wilczek:1978xi,Ong:1978tq,Davidson:1979wt,Chakrabarti:1979vy,Yanagida:1979gs,Yanagida:1980xy,Berezhiani:1990wn,Nagoshi:1990wk}, is to work with what we already have at disposal and gauge directly the flavor (family) index of the standard (i.e., observed) fermions. Assuming that there are three families of standard fermions, we obtain this way the \emph{gauge flavor dynamics} (g.f.d.) with the gauge group $\group{SU}(3)_{\mathrm{F}}$ and with the corresponding coupling constant $h$ being thus the only free parameter of the theory. The new flavor dynamics is assumed to be responsible for the EWSB. The structure of the fermion gauge representations ensures that the model is \emph{chiral}. The aim of this paper is to present the model in more detail and to provide some new arguments in its favor.

Before we turn into technical details in the following sections, let us first sketch the main points of the presented scheme:
\begin{itemize}
  \item Since the postulated flavor symmetry $\group{SU}(3)_{\mathrm{F}}$ is not a symmetry of the observed fermion mass spectrum, it has to be spontaneously broken along the mass generation in infrared. Therefore we demand that the g.f.d.~is asymptotically free and assume that it completely self-breaks bellow some scale $\Lambda_{\mathrm{F}}$.
  \item Because in QCD we trust, the g.f.d.~must \emph{not} be vector-like. For if it were, it would be confining at momenta bellow $\Lambda_{\mathrm{F}}$ in contradiction with experiment.
  \item Assumed spontaneous breaking of the $\group{SU}(3)_{\mathrm{F}}$ means generation of masses of the eight \emph{flavor-gluons}, the $\group{SU}(3)_{\mathrm{F}}$ gauge bosons. Phenomenologically, since the exchanges of flavor-gluons generally change flavor, they have to be very heavy, with masses to be at least of the order of $1\,000\,\mathrm{TeV}$ \cite{Eichten:1979ah}.
  \item Non-perturbative flavor-gluon exchanges between the left-handed and the right-handed fermion fields induced by the large effective sliding flavor charge $\bar h(q^2)$ lead to generation of fermion masses. The mass differences among the fermions of different electric charges are guaranteed by assignments of their chiral components in specific representations (triplets or antitriplets) of $\group{SU}(3)_{\mathrm{F}}$. Neutrinos turn out to be Majorana particles.
  \item All fermion masses are given in terms of the large flavor-gluon masses. We associate their smallness with the critical scaling \cite{Miransky:1996pd,Braun:2010qs} which occurs not earlier than near a non-trivial non-perturbative infrared (IR) fixed point.
  \item The fermion masses, generated dynamically in the course of breaking the $\group{SU}(3)_{\mathrm{F}}$ symmetry, break also the electroweak symmetry $\group{SU}(2)_{\mathrm{L}} \times \group{U}(1)_{\mathrm{Y}}$ down to $\group{U}(1)_{\mathrm{em}}$. In other words, the EWSB is a consequence of the flavor-symmetry breaking. The resulting masses of $W^\pm$ and $Z$ are thus expressed in terms of the fermion masses (or more precisely, in terms of the fermion self-energies). This means that the electroweak scale is not genuinely inherent in the g.f.d.~model, rather it is given by the top-quark mass.
\end{itemize}

In principle the ultimate test of the model is simple. Compute its mass spectrum and compare it with the observed one. Because the model has only one free parameter, the mass ratios should be uniquely given and the model is either right or plainly wrong. In reality the situation is anything but simple: (i) For asymptotically free quantum field theories, which are strongly coupled in the infrared, the reliable non-perturbative computations of the spectrum in the continuum are not available. (ii) Even though the model deals with the chiral fermion fields which are experimentally observed, it is not known how to put them on the lattice \cite{Creutz:2004eq,Kaplan:2009yg}. (iii) Experimental tests of asymptotic freedom in our model, so famous in QCD \cite{[{For a review, see }][{}]Bethke:2006ac}, are disqualified by the enforced extreme heaviness of the flavor-gluons: The flavor-gluon interactions between leptons and quarks are effectively extremely weak up to very high-momentum transfer.

We believe that the model should nevertheless manifest itself experimentally at accessible energies without knowledge of the spectrum, basically due to its rigidity. It has a restricted form of the neutrino sector and is characterized by definite global symmetries. The global symmetries being spontaneously broken manifest themselves by a definite spectrum of composite scalars, (pseudo-)Nambu--Goldstone (NG) bosons. We list both characteristic properties of the model with the aim of suggesting their experimental signatures.

The paper is organized as follows: In Sec.~\ref{sec:lagrangian} we define the model in terms of the fermion flavor representations and the corresponding Lagrangian. In Sec.~\ref{sec:global_sym} we analyze the global symmetries with focus on their anomalies. Sec.~\ref{sec:fla_sym_bre} is devoted to the very argumentation in favor of the assumed spontaneous breaking of the flavor symmetry. In Sec.~\ref{sec:fer_mass_gen} we discuss generation of the fermion masses, whose effect on EWSB and the $W^\pm$ and $Z$ boson masses is investigated in the subsequent Sec.~\ref{sec:EW_sym_bre}. In Sec.~\ref{sec:experimental} some experimental consequences, mainly the spectrum of various (pseudo-)Nambu--Goldstone bosons, are discussed. This is followed by comparison of the present model with other models in Sec.~\ref{sec:comparison} and we conclude in Sec.~\ref{sec:conclusions}.

\section{The Lagrangian}
\label{sec:lagrangian}

\begin{table}[t]
\begin{tabular}{l|ccccc|cc|c}
& $q_L$ & $u_R$ & $d_R$ & $\ell_L$ & $e_R$    & $A_{\mathrm{quarks}}$ & $A_{\mathrm{EW-leptons}}$ & $A_{\nu_R}$ \\
\hline
\hline
case I  & $\mathbf{3}$ & $\mathbf{3}$ & $\overline{\mathbf{3}}$ & $\overline{\mathbf{3}}$ & $\mathbf{3}$    & $-6$ & $2+1$ & $3$ \\
case II & $\mathbf{3}$ & $\mathbf{3}$ & $\overline{\mathbf{3}}$ & $\overline{\mathbf{3}}$ & $\overline{\mathbf{3}}$    & $-6$ & $2-1$ & $5$ \\
\hline
\hline
\end{tabular}
\caption{\small The flavor representation settings for the cases I and II, together with the sum of anomaly coefficients \eqref{anomalyC} for quarks $A_{\mathrm{quarks}}$, for electroweakly charged leptons $A_{\mathrm{EW-leptons}}$ and for right-handed neutrinos $A_{\nu_R}$, which must sum up to zero.}
\label{f_setting}
\end{table}

\subsection{Standard fermions}

The fermion content consists, apart from the right-handed neutrinos to be discussed subsequently, of the experimentally established \emph{three} electroweakly $\group{SU}(2)_\mathrm{L}\times \group{U}(1)_\mathrm{Y}$ and color $\group{SU}(3)_\mathrm{C}$ identical families of the standard chiral quark and lepton fields $q_{iL}=(u_{iL}, d_{iL})^{\T}$, $u_{iR}$, $d_{iR}$, $\ell_{iL}=(\nu_{iL}, e_{iL})^{\T}$, $e_{iR}$, with $i=1,2,3$.

We gauge the flavor index $i$ in such a way that for different electric charges the corresponding mass matrices (or, in our treatment, the proper self-energies) should come out different. Therefore we put the chiral fermion fields into $\group{SU}(3)_\mathrm{F}$ triplet/antitriplet representations as follows\footnote{Similar setting is used in \cite{Appelquist:2003hn}.}(see also Tab.~\ref{f_setting}):

(I) The choice of $q_L$ as a triplet $(\trip)$ (i.e., both $u_L$ and $d_L$ are triplets) is mere convention. Then, to distinguish $u_R$ and $d_R$ we must set one to a triplet and the other to an antitriplet. We choose without loss of generality $(u_R,d_R)=(\trip,\atrip)$.

(II) It follows that $\ell_L$ cannot be a triplet. For if it were, the charged lepton mass matrix would be equal either to the $u$-type or the $d$-type quark mass matrix. Hence, $\ell_L$ (i.e., both $\nu_L$ and $e_L$) must be an antitriplet $(\atrip)$.

(III) It then follows that $e_R$ can be either triplet $(\trip)$ or antitriplet $(\atrip)$. We refer to the former possibility as the \emph{case I} and the latter as the \emph{case II}.

\subsection{Right-handed neutrinos}

In order to account for the right-handed neutrinos, we impose at this point two important quantum field theoretical restrictions:

First, a gauge theory containing chiral fermion fields must be free of gauge axial anomalies in order to be well defined \cite{Gross:1972pv}. Simple inspection reveals that the present model with the fermion content introduced so far is not. To make it anomaly free we take the liberty of introducing the right-handed neutrino fields, the electroweak and color singlets, in appropriate $\group{SU}(3)_\mathrm{F}$ representations. `Minimal' anomaly-free solutions in cases I or II amount to introducing three or five right-handed neutrino flavor triplets $\nu_{R}^s$, $s=1, 2, 3$ or $s=1,\dots, 5$, respectively. `Nonminimal' solutions also exist (see Appendix \ref{appendix}).

Second, to have a chance to generate masses dynamically the model must not be infrared stable at the origin \cite{Stern:1976jg}. Therefore our model must stay asymptotically free, what is fulfilled only for certain combinations of right-handed neutrino flavor multiplets of rather lower number and dimension.

There is only a limited number of physically viable solutions fulfilling both restrictions (see Tab.~\ref{AAfreeSettings}).

\subsection{Minimal solution of case I}

From now on we will consider for definiteness the minimal solution of the case I, i.e., with $e_R$ being in \emph{triplet} and with \emph{three} right-handed neutrino triplets $\nu_R^s$, $s=1, 2, 3$. The perturbative one-loop beta function then has the form \eqref{beta},
\begin{eqnarray}
\beta(h) &=& -\frac{h^3}{16\pi^2}\big[11-\tfrac{1}{3}N^{\mathrm{ew}}-\tfrac{1}{3}N^{\nu_R}\big] \,,
\end{eqnarray}
with $N^{\mathrm{ew}}=15$ and $N^{\nu_R}=3$, and $h$ being the flavor gauge coupling parameter. Hence, the model is asymptotically free at short distances and is perturbatively well defined. In particular, the momentum dependent sliding coupling $\bar h^2(q^2)$ at large momenta is
\begin{eqnarray}
\frac{\bar h^2(q^2)}{4\pi} &=& \frac{12 \pi}{(33-N^{\mathrm{ew}}-N^{\nu_R}) \ln (q^2/\Lambda^2_{\mathrm{F}})}
\,.
\label{hPT}
\end{eqnarray}
Clearly, $\Lambda_{\mathrm{F}}$, the energy scale where the flavor dynamics becomes strongly coupled, is theoretically arbitrary and should be fixed from an experiment.

To summarize, the dynamics which intends to replace the Higgs sector of the SM is thus defined by the non-vector-like Lagrangian
\begin{eqnarray}
\label{gfdL}
{\cal L}_{\mathrm{g.f.d.}} &=& \phantom{+}
\bar q_L \I \slashed{\D} q_L + \bar u_R \I \slashed{\D} u_R + \bar d_R \I \slashed{\D} d_R
\nonumber \\ &&
+ \bar \ell_L \I \slashed{\D} \ell_L + \bar e_R \I \slashed{\D} e_R + \bar \nu_{R}^s \I \slashed{\D} \nu_{R}^s
-\frac{1}{4}F_{a\mu\nu}F_a^{\mu\nu} \,.
\nonumber \\ &&
\end{eqnarray}
Covariant derivatives $\D^\mu=\partial^\mu+\im hC_{a}^\mu T_{f_{L/R},a}$ of the respective fermions $f_L$, $f_R$ contain either triplet $T_{f_{L/R},a}(\trip)=\tfrac{1}{2}\lambda_a$ or antitriplet $T_{f_{L/R},a}(\atrip)=-\tfrac{1}{2}\lambda_a^*$ generators of $\group{SU}(3)_\mathrm{F}$ according to the case I fermion assignment defined above. Summation is assumed over the neutrino index $s=1,2,3$. The $F^{\mu\nu}_a=\partial^\mu C_{a}^\nu-\partial^\nu C_{a}^\mu+hf_{abc}C_{b}^\mu C_{c}^\nu$ is the $\group{SU}(3)_\mathrm{F}$ flavor-gluon field strength tensor.

Complete $\group{SU}(2)_\mathrm{L} \times \group{U}(1)_\mathrm{Y} \times \group{SU}(3)_\mathrm{C} \times \group{SU}(3)_\mathrm{F}$ gauge
invariant Lagrangian ${\cal L}$ of the world, including QCD and
electroweak interactions, is obtained from ${\cal L}_{\mathrm{g.f.d.}}$ by
gauging it with respect to $\group{SU}(2)_\mathrm{L} \times \group{U}(1)_\mathrm{Y} \times \group{SU}(3)_\mathrm{C}$ in the usual way. We keep the standard
abbreviation: The electroweak gauge dynamics
is characterized by the coupling constants $g$, $g'$ and by the field
strength tensors $W_a^{\mu\nu}$, $Y^{\mu\nu}$, respectively; the QCD gauge
dynamics is characterized by the coupling constant $g_\mathrm{s}$ and by the
field strength tensor $G_a^{\mu\nu}$.

\section{Global symmetries}
\label{sec:global_sym}

Apart from the gauge symmetries, the complete classical Lagrangian $\eL$ possesses also a rich spectrum of global symmetries.

First, the gauge symmetries do not distinguish among the three new triplets of right-handed neutrinos $\nu_{R}^s$. It is a new chiral global non-Abelian symmetry that rotates them. We call it the \emph{sterility symmetry} $\group{SU}(3)_\mathrm{S}$. Its current is
\begin{eqnarray}\label{Eq:ster:curr:nA}
J^{\mu,\sigma}_\mathrm{S} & = & \bar \nu_{R}^s\left[\tfrac{1}{2}\lambda^\sigma\right]^{sr}\gamma^{\mu}\,\nu_{R}^r \,,
\end{eqnarray}
where index $\sigma$ labels eight $\group{SU}(3)_\mathrm{S}$ generators given by the
Gell-Mann matrices $\tfrac{1}{2}\lambda^\sigma$.

Second, both the gauge and the global non-Abelian symmetries tie together different chiral fermion fields
and leave the room for only six Abelian symmetries corresponding to common
phases of the fields $\ell_L$, $\nu_{R}^s$, $e_R$, $q_L$, $u_R$, $d_R$.
One combination of chiral currents defines the gauged weak hypercharge with the convention $Y(\ell_L,\nu_{R}^s,e_R,q_L,u_R,d_R)=(-1,0,-2,\tfrac{1}{3},\tfrac{4}{3},-\tfrac{2}{3})$.

In analogy with the SM we define the remaining five global Abelian symmetries
\begin{subequations}
\begin{eqnarray}
J^{\mu}_\mathrm{B} &=& \frac{1}{3}[\bar q_L \gamma^{\mu}q_L+ \bar u_R\gamma^{\mu}u_R+\bar d_R \gamma^{\mu}d_R] \,,
\label{jB} \\
J^{\mu}_{\mathrm{B}_{5}} & = & \frac{1}{3}[-\bar q_L \gamma^{\mu}q_L+ \bar u_R\gamma^{\mu}u_R+ \bar d_R \gamma^{\mu}d_R] \,,
\label{jB5} \\
J^{\mu}_\mathrm{L} & = & \bar \ell_L \gamma^{\mu}\ell_L+\bar e_R\gamma^{\mu}e_R \,,
\label{jL} \\
J^{\mu}_{\mathrm{L}_{5}} & = & {-\bar \ell_L \gamma^{\mu}\ell_L+\bar e_R\gamma^{\mu}e_R} \,,
\label{jL5} \\
J^{\mu}_\mathrm{S} & = & \frac{1}{3}\bar \nu_{R}^s\gamma^{\mu}\nu_{R}^s \,.
\label{jS}
\end{eqnarray}
\end{subequations}

In the following we compute the chiral anomalies of the global chiral currents at one-loop to check the status of the corresponding global symmetries at the quantum level \cite{'tHooft:1976up}. Straightforward computation of the underlying anomalous triangles \cite{Adler:1969gk,Bell:1969ts} reveals that the non-Abelian sterility symmetry $\group{SU}(3)_\mathrm{S}$ is not anomalous, i.e.,
\begin{eqnarray}\label{Eq:div:ster:nA}
\partial_\mu J^{\mu,\sigma}_\mathrm{S} & = & 0 \,.
\end{eqnarray}
On the other hand it results in nonzero anomalous divergences of the five Abelian currents:

\begin{subequations}
\begin{eqnarray}
\partial_{\mu}J^{\mu}_\mathrm{B} & = & 3\frac{{g'}^{2}}{8\pi^2}Y\tilde Y -3\frac{g^2}{32\pi^2}W\tilde W \,,
\\
\partial_{\mu}J^{\mu}_\mathrm{L} & = & 3\frac{{g'}^{2}}{8\pi^2}Y\tilde Y - 3\frac{g^2}{32\pi^2}W\tilde W - \frac{h^2}{32\pi^2}F\tilde F \,,
\\
\partial_{\mu}J^{\mu}_{\mathrm{B}_{5}} & = & \frac{11}{3}\frac{{g'}^{2}}{8\pi^2}Y\tilde Y + 3\frac{g^2}{32\pi^2}W\tilde W + 4\frac{h^2}{32\pi^2}F\tilde F
\nonumber \\ &&
+ 4\frac{g_\mathrm{s}^2}{32\pi^2}G\tilde G \,,
\\
\partial_{\mu}J^{\mu}_{\mathrm{L}_{5}} & = & 9\frac{{g'}^{2}}{8\pi^2}Y\tilde Y + 3\frac{g^2}{32\pi^2}W\tilde W + 3\frac{h^2}{32\pi^2}F\tilde F \,, \qquad
\\
\partial_{\mu}J^{\mu}_{\mathrm{S}} & = & \frac{h^2}{32\pi^2}F\tilde F \,,
\end{eqnarray}
\end{subequations}
where for all field strength tensors $X=Y, W, G, F$ their duals are
defined as $\tilde X_{\alpha \beta}=\tfrac{1}{2}\epsilon_{\alpha \beta \rho \sigma}X^{\rho \sigma}$.

Conclusions of this computation are standard:

(I) Out of five classically conserved currents only one of their linear
combinations corresponds to the true symmetry at quantum level. It is the current $J^{\mu}_{\mathrm{B}-\mathrm{L}-\mathrm{S}}$, the straightforward analog of the SM current $J^{\mu}_{\mathrm{B}-\mathrm{L}_{\mathrm{SM}}}$, since the extended lepton number $L+S$ has the same anomaly as the SM lepton number $L_{\mathrm{SM}}$.

(II) The baryon number current $J^{\mu}_{\mathrm{B}}$ is the only global current that will not be affected by the dynamical symmetry breaking. Nevertheless, like in the SM, it is broken by the anomaly. Because the anomaly is given merely by the electroweak dynamics, it is negligible and the baryon number $B$ is a rather good approximate symmetry \cite{Peccei:1998jv}. Due to the electroweak anomaly, the baryon phase transformation rotates the $\group{SU}(2)_\mathrm{L}$ weak CP violating $\theta$-term out of the Lagrangian, and thus makes the corresponding parameter unobservable \cite{Krasnikov:1978dg,Anselm:1993uj}.

(III) Out of the remaining currents another linear combination can be constructed, which corresponds again to rather good symmetry, broken only by electroweak anomaly. Example of such a current is $J^{\mu}_{\mathrm{L}_{5}+3\mathrm{L}}$.

(IV) The current broken by the QCD anomaly does not correspond to any
symmetry at all as the effect of the strong and topologically non-trivial QCD dynamics is not negligible. However phenomenologically, its presence is welcome as it provides the Peccei--Quinn transformation \cite{Peccei:1977hh,*Peccei:1977ur}. Example of such a current is $J^{\mu}_{\mathrm{B}_{5}-4\mathrm{S}}$.

(V) The anomaly given by g.f.d.~breaks heavily the remaining symmetry corresponding to the current, e.g., $J^{\mu}_{\mathrm{S}}$.

Important question to ask is what happens to the global Abelian
symmetry or to the `would-be' symmetries when nonzero fermion masses
are spontaneously generated. We answer it in Sec.~\ref{sec:experimental}.

\section{Flavor symmetry self-breaking}
\label{sec:fla_sym_bre}

The flavor symmetry is not a property of the fermion mass spectrum, therefore it has to be dynamically broken. We do not introduce any other dynamics in order to provide the breaking, but we assume that the g.f.d.~self-breaks \cite{Eichten:1974et}. In this section we first present some general aspects of such symmetry breakdown and introduce this way also the notation to be used in the following sections. Then we give some more physical view of the way the flavor symmetry is assumed to be broken.

\subsection{Masses as order parameters}
\label{subsec:ord_par}

At the Lagrangian level the masslessness of fermion fields is perturbatively protected by chiral symmetries, the flavor and electroweak symmetries in particular. The masslessness of gauge fields is protected by the gauge nature of the symmetries. \emph{Massless fields can, however, excite massive particles}, if the protective symmetries are spontaneously broken.

\subsubsection{Fermions}
\label{ssec:fermions}

Massless fermion fields excite massive fermions if the ground state is not invariant under independent rotations of their left-handed and right-handed components \cite{Nambu:1961tp}.

Operationally this manifests by nonzero chirality-changing parts $\mathbf{\Sigma}_f$ of the full propagators $\I S_f = \langle f \bar f \rangle$, $f=n,e,u,d$. The field $f$ is for the charged fermions defined in terms of the original chiral fields simply as
\begin{eqnarray}
\label{f=ude}
f &\equiv& f_L + f_R \quad\quad (f=u,d,e) \,,
\end{eqnarray}
whereas for neutrinos it is more convenient to make use of fermion charge conjugation and to define the Nambu--Gorkov doublet
\begin{eqnarray}
\label{f=n}
n &\equiv&
\left(\begin{array}{c} \nu_L + (\nu_L)^\C \\ \nu_R^1 + (\nu_R^1)^\C \\ \nu_R^2 + (\nu_R^2)^\C \\ \nu_R^3 + (\nu_R^3)^\C \end{array}\right) \,.
\end{eqnarray}
The corresponding propagators $S_f$ are considered for the sake of simplicity of the special form \cite{Benes:2009iz}
\begin{eqnarray}
\label{sigma_ansatz}
S_f^{-1}(p) &=& \slashed{p}-\mathbf{\Sigma}_f(p^2) \,,
\end{eqnarray}
with
\begin{eqnarray}
\mathbf{\Sigma}_f &=& \Sigma_f P_L+\Sigma_f^{\dag}P_R
\end{eqnarray}
and $P_{L,R}=\tfrac{1}{2}(1\mp\gamma_5)$. Notice that $\Sigma_f$ are in principle arbitrary complex $p^2$-dependent matrices of the dimension $3 \times 3$ for charged fermions and $12 \times 12$ for neutrinos. Moreover, the neutrino matrix $\Sigma_n$ is symmetric. The inverse of \eqref{sigma_ansatz} is explicitly given by
\begin{eqnarray}
\label{S}
S_f(p) &=& \phantom{+}\,
(\slashed{p}+\Sigma_f^{\dag})(p^2-\Sigma_f\Sigma_f^{\dag})^{-1}P_L
\nonumber \\ &&
+\,
(\slashed{p}+\Sigma_f)(p^2-\Sigma_f^{\dag}\Sigma_f)^{-1}P_R \,.
\end{eqnarray}
The fermion mass spectrum is then given by the poles of the full propagator \eqref{S}, i.e., by the solutions of the equation
\begin{eqnarray}
\det\big[p^2-\Sigma_f(p^2)\Sigma_f^{\dag}(p^2)\big] &=& 0 \,.
\label{poles_fermion}
\end{eqnarray}

Breaking of the $\group{SU}(3)_{\mathrm{F}}$ symmetry by the fermion self-energies can be written compactly as
\begin{eqnarray}
\mathbf{\Sigma}_f T_{f,a} - \bar T_{f,a} \mathbf{\Sigma}_f &\neq& 0 \,,
\end{eqnarray}
where the $\group{SU}(3)_{\mathrm{F}}$ generators $T_{f,a}$ in the bases \eqref{f=ude}, \eqref{f=n} are given as
\begin{eqnarray}
T_{f,a} &\equiv& T_{f_L,a} P_L + T_{f_R,a} P_R \quad\quad (f=u,d,e) \,, \\
T_{n,a} &\equiv& \phantom{+}\,
\left(\begin{array}{cccc}
T_{\nu_L,a} & 0 & 0 & 0 \\
0 & - T_{\nu_R^1,a}^\T & 0 & 0 \\
0 & 0 & - T_{\nu_R^2,a}^\T & 0 \\
0 & 0 & 0 & - T_{\nu_R^3,a}^\T \end{array}\right) P_L
\nonumber \\ && +\,
\left(\begin{array}{cccc}
- T_{\nu_L,a}^\T & 0 & 0 & 0 \\
0 & T_{\nu_R^1,a} & 0 & 0 \\
0 & 0 & T_{\nu_R^2,a} & 0 \\
0 & 0 & 0 & T_{\nu_R^3,a} \end{array}\right) P_R
\end{eqnarray}
and where $\bar T_{f,a} \equiv \gamma_0 T_{f,a} \gamma_0$.

For the sake of later references we introduce the $\group{SU}(3)_{\mathrm{F}}$ generators also for the electroweak doublets $\mathcal{D}_L = q_L, \ell_L$. Taking into account their general structure $\mathcal{D}_L = (U_L,D_L)^\T$ (with $U_L=u_L,\nu_L$ and $D_L=d_L,e_L$) and $T_{U_L,a} = T_{D_L,a}$, we find
\begin{eqnarray}
T_{\mathcal{D}_L,a} &=&
\left(\begin{array}{cc} 1 & 0 \\ 0 & 1 \end{array}\right) T_{U_L,a} =
\left(\begin{array}{cc} 1 & 0 \\ 0 & 1 \end{array}\right) T_{D_L,a} \,,\qquad
\end{eqnarray}
with the unit matrix operating in the electroweak doublet space.

\subsubsection{Flavor-gluons}

Massless gauge fields excite massive vector particles if the ground state is not invariant under global symmetry underlying the gauge one \cite{Englert:1964et,Higgs:1964pj,Guralnik:1964eu}. The longitudinal polarization state of such a massive vector particle emerges as the `would-be' NG boson of the broken symmetry. Operationally it manifests by massless pole in the transverse polarization tensor
\begin{eqnarray}
\Pi^{\mu\nu}_{ab}(q) &\equiv& (q^2g^{\mu\nu}-q^{\mu}q^{\nu})\Pi_{ab}(q^2) \,,
\label{Pimn}
\end{eqnarray}
i.e., by the $\Pi_{ab}$ of the form
\begin{eqnarray}
\Pi_{ab}(q^2) &=& \frac{1}{q^2} M_{ab}^2(q^2) \,,
\label{PiIR}
\end{eqnarray}
where $M_{ab}^2(q^2)$ is a momentum-dependent symmetric $8 \times 8$ matrix, \emph{regular} at $q^2=0$. Massiveness of the flavor-gluons is then visible from their full propagator $\I \Delta^{\mu\nu}_{ab} = \langle C_a^{\mu} C_b^{\nu} \rangle$, having the form
\begin{eqnarray}
\Delta^{\mu\nu}_{ab}(q) &=&
-\frac{1}{q^2}
\bigg(g^{\mu\nu}-\frac{q^{\mu}q^{\nu}}{q^2}\bigg)\Big[\big(1-\Pi(q^2)\big)^{-1}\Big]_{ab}\quad
\label{Delta}
\end{eqnarray}
in the Landau gauge. Poles of this full propagator $\Delta^{\mu\nu}_{ab}$ are given by the equation
\begin{eqnarray}
\det\big[q^2-M^2(q^2)\big] &=& 0 \,,
\label{M}
\end{eqnarray}
solutions of which define the flavor-gluon mass spectrum.

The flavor-gluon polarization tensor $\Pi_{ab}$ breaks the $\group{SU}(3)_{\mathrm{F}}$ symmetry once
\begin{eqnarray}
[T_a,\Pi] &\neq& 0 \,,
\end{eqnarray}
where $T_a$ are the $\group{SU}(3)_{\mathrm{F}}$ generators in the adjoint representation and $\Pi$ is the polarization tensor with suppressed indices.

\begin{center}*\end{center}

\begin{center}
\begin{figure}[t]
\begin{center}
\includegraphics[width=0.5\textwidth]{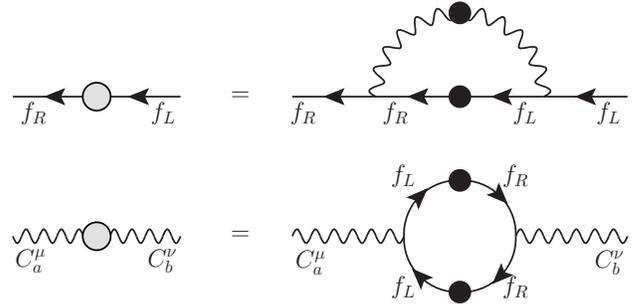}
\caption[]{Mutual interactions of the flavor-gluons and the fermions, giving rise to the $\group{SU}(3)_{\mathrm{F}}$ symmetry-breaking fermion self-energies $\Sigma_f$ and flavor-gluon polarization tensor $\Pi_{ab}^{\mu\nu}$. Contributions of the pure gauge diagrams are omitted in the second line. The black and grey blobs stand for the full and 1PI propagators, respectively.}
\label{fig:SDE}
\end{center}
\end{figure}
\end{center}


\emph{The symmetry-breaking $\Sigma$s and flavor-gluon $\Pi$ are thus the basic lowest-dimensional order parameters.}

The lesson learned from QCD is that the global $\group{SU}(3)_\mathrm{F}$ symmetry can hardly be broken by the underlying gauge dynamics, without (non-vector-like) fermions \cite{Vafa:1983tf,Vafa:1984xg}. Thus, in order to generate non-vanishing order parameters (i.e., in our case the flavor symmetry-breaking fermion and flavor-gluon propagators) mutual interactions of both the fermions and the gauge bosons have to be taken into account. In physical terms the dynamical flavor-gluon mass generation means \cite{Eichten:1974et,Smit:1974je,Binosi:2009qm} that the flavor-gluon exchanges between themselves and between the fermions (including neutrinos) of both chiralities self-consistently provide strong attraction necessary for the formation of eight `would-be' NG bosons composed both from the flavor-gluons and the fermions. They express themselves as the above discussed massless pole of the flavor-gluon polarization tensor. Schematically and most straightforwardly, one can imagine for that purpose the diagrams depicted in Fig.~\ref{fig:SDE}. They are subset of the full tower of integral Schwinger--Dyson (SD) equations for all Green's functions of the theory, truncated in our case at the level of three-point functions.

\subsection{Physical view}
\label{Physical_view}

The more formal approach based on the SD equations will be utilized to some extend in the following sections. It has, however, the disadvantage of giving little physical insight on what is actually going on in the course of the flavor symmetry breaking. Let us now spend few words on this issue, without going too much into detail. Technically the simplest physical description of the flavor symmetry breaking in the g.f.d.~model, used in a similar context also in \cite{Nagoshi:1990wk}, can be pursued along the following line:

In the pure $\group{SU}(3)$ gauge dynamics it is conceivable that one common gauge boson mass parameter appears. It accompanies the massless pole in the gauge boson proper self-energy \cite{Cornwall:1973ts,Binosi:2009qm} which is a result of a strong coupling. In the g.f.d.~model it can be effectively described by the mass term
\begin{equation}
M_{0}^2C_{a\mu}C_{a}^{\mu}
\end{equation}
leaving the global $\group{SU}(3)_\mathrm{F}$ unbroken.

Massive flavor-gluon exchanges between left- and right-handed fermion fields in triplets and antitriplets yield effective four-fermion
interactions
\begin{eqnarray}
{\cal L}_{f^4} &=& \frac{h^2}{M_{0}^2}j_a^{\mu}j_{a\mu} \,,
\end{eqnarray}
where the electroweakly invariant current reads
\begin{eqnarray}
j_a^{\mu} &=& \sum_{\mathcal{D}_L} \mathcal{\bar D}_{L}\gamma^\mu T_{\mathcal{D}_L,a}\mathcal{D}_{L} + \sum_{f_R} \bar f_{R}\gamma^\mu T_{f_R,a}f_{R} \,.
\end{eqnarray}
The sums run over the left-handed doublets $\mathcal{D}_L=q_L,\ell_L$ and the right-handed singlets $f_R=u_R,d_R,e_R,\nu_{R}^s$.

\subsubsection{Dirac masses}

The four-fermion interactions, upon Fierz rearrangements into the form $(\mathcal{\bar D}_L f_R)(\bar f_R \mathcal{D}_L)$, provide attractive channels inducing the formation of effective composite Higgs fields which are electroweak doublets and color singlets. The effective Higgs fields are interpolated by fermion field bilinears
\begin{eqnarray}
\Phi^{(u)} & \propto & \bar{u}_R q_L \,, \\
\Phi^{(d)}_{A} & \propto & \bar{d}_R T_A q_L \,, \\
\Phi^{(e)}_{A} & \propto & \bar{e}_R T_A \ell_L \,, \\
\Phi^{(\nu)s}_{A} & \propto & \bar{\nu}^{s}_R T_A \ell_L \,,
\end{eqnarray}
where the index $A=2,5,7$ labels the anti-symmetric generators and indicates the flavor triplet representation.
The vacuum expectation values of their electrically neutral components are related to the fermion condensates as
\begin{eqnarray}
\langle\Phi^{f}\rangle & = & \frac{1}{M_{0}^2}\langle\bar{f}_R f_L\rangle
\end{eqnarray}
and they result from a minimization of effective potential for the Higgs fields $V(\Phi)$ which is induced by the strong g.f.d. We do not specify here the potential. The four-fermion interactions after the introduction of effective Higgs fields yield the Yukawa interactions \cite{Hosek:1982cz,Hosek:1985jr,Miransky:1988xi,Bardeen:1989ds}.
\begin{eqnarray}
{\cal L}^{\mathrm{Yukawa}}_{\mathrm{Dirac}} & = &
-8h^2\Phi^{(u)\dag}\bar{u}_R q_L-\frac{8}{3}h^2\Phi^{(d)\dag}_{A}\bar{d}_R T_A q_L \\
& & -\frac{8}{3}h^2\Phi^{(e)\dag}_{A}\bar{e}_R T_A \ell_L -\frac{8}{3}h^2\Phi^{(\nu)s\dag}_{A}\bar{\nu}^{s}_R T_A \ell_L +\mathrm{h.c.} \nonumber
\end{eqnarray}
After the Higgs condensation Dirac masses for fermions are generated. Analogously the four-fermion interactions yield mass terms for the effective Higgs fields.
\begin{eqnarray}
{\cal L}_{\Phi^\dag\Phi} & = &
-8h^2M_{0}^2\Phi^{(u)\dag}\Phi^{(u)}-\frac{8}{3}h^2 M_{0}^2\sum_{f=d,e,\nu}\Phi^{(f)\dag}_{A}\Phi^{(f)}_{A} \,,
\nonumber \\ &&
\end{eqnarray}
where we suppressed summation over electroweak and sterility indices.

\subsubsection{Majorana masses}

For neutrinos there are moreover channels of the form $(\bar{f} C\bar{f}^\T)(f^\T C f)$, where $C$ is the matrix of charge conjugation, providing their Majorana masses. Introducing effective composite Higgs fields for the attractive Majorana channels
\begin{eqnarray}
\Upsilon^{(\nu)}_A & \propto & \bar{\ell}_L T_A C\bar{\ell}^{\T}_L \,, \\
\Upsilon^{(\nu)sr}_{A} & \propto & \bar{\nu}^{s}_R T_A C\bar{\nu}^{r\T}_R \,,
\end{eqnarray}
the former $\Upsilon$ Higgs field is flavor and electroweak triplet, while the latter are flavor triplets and electroweak singlets. The relevant Yukawa interactions and Higgs field mass terms are
\begin{eqnarray}
{\cal L}^{\mathrm{Yukawa}}_{\mathrm{Majorana}}
& = &
-\frac{8}{3}h^2 \nu^{\T}_L \Upsilon^{(\nu)}_A T_A C\nu_L \nonumber \\
& & -\frac{8}{3}h^2\Upsilon^{(\nu)sr}_{A}\nu^{r\T}_R T_A C\nu^{s}_R +\mathrm{h.c.} \\
{\cal L}_{\Upsilon\Upsilon}
& = &
-\frac{8}{3}h^2 M_{0}^2 \Upsilon^{(\nu)}_A \Upsilon^{(\nu)}_A-\frac{8}{3}h^2 M_{0}^2 \Upsilon^{(\nu)sr}_A \Upsilon^{(\nu)sr}_A \,.
\nonumber \\ &&
\end{eqnarray}

\begin{center}*\end{center}

The number of vacuum expectation values of composite triplet Higgs fields $\langle\Phi_{A}\rangle$ and $\langle\Upsilon_A\rangle$ is more than sufficient to break the gauge $\group{SU}(3)_\mathrm{F}$ symmetry completely.

Three linear combinations of components of the electroweak doublets $\Phi_A$ are the `would-be' NG bosons, the longitudinal components of massive electroweak bosons.

Presumably the mass $M_0$ is of the order of $\Lambda_\mathrm{F}$. The vacuum expectation value $\langle\Phi^{(u)}\rangle$ provides the magnitude of the top-guark mass, and thus the electroweak scale $\Lambda_{\mathrm{EW}}$. The global flavor symmetry does not protect smallness of neither $M_0$ nor $\langle\Phi^{(u)}\rangle$. The phenomenologically necessary smallness of ratio
\begin{eqnarray}
\frac{\Lambda_{\mathrm{EW}}}{\Lambda_\mathrm{F}}\approx\frac{\langle\Phi^{(u)}\rangle}{M_0} &\ll& 1
\end{eqnarray}
has to result from the critical scaling. The situation is different for the other mass parameters. On top of their suppression by the critical scaling there is also the global $\group{SU}(3)_\mathrm{F}$ which protects their smallness. From this point of view the non-vector-like presence of fermions is indispensable for the g.f.d.~self-breaking scenario.

\subsection{Effective flavor charge}

At asymptotically large momenta where the flavor symmetry is not broken and the g.f.d.~is in perturbative r\'{e}gime, the flavor-gluon self-energy has the form $\Pi_{ab}=\delta_{ab}\Pi_\mathrm{pert.}$, and the effective momentum-dependent coupling constant is
defined as \cite{Binosi:2009qm}
\begin{eqnarray}
\frac{\bar h^2_{ab}(q^2)}{q^2} &=& \delta_{ab}\frac{h^2}{q^2\big(1-\Pi_\mathrm{pert.}(q^2)\big)} \,. \label{barh}
\end{eqnarray}
That is, the behavior of $\bar h^2_{ab}(q^2)/(4\pi)$ is perturbative and has the form \eqref{hPT}. This is important for knowing how the
solutions $\Sigma$ of the SD equations go to zero at high momenta
\cite{Pagels:1978ba}:
\begin{eqnarray}
\Sigma(p^2) &\rightarrow& \frac{1}{p^2}\ln^{\gamma-1}(p^2/\Lambda^2_{\mathrm{F}}) \,,
\end{eqnarray}
where $\gamma=12/(33-N^{\mathrm{ew}}-N^{\nu_R})$.

The fermion masses are, however, determined by the behavior of
$\Sigma$s at low momenta, at which $\bar h_{ab}^2(q^2)$ is expected
large and is entirely unknown. Important is the following: Using of
the low-momentum $\Pi$, (\ref{PiIR}), in (\ref{barh}) yields correctly
the massive flavor-gluon propagator, but erroneous (vanishing) $\bar
h^2_{ab}(q^2)$ \cite{Binosi:2009qm} at low momenta. Consequently, the formula
(\ref{barh}) should be modified. Physically this is not surprising.
At low momenta the strongly coupled g.f.d.~is expected to
produce bound states, and if some of them effectively interact with
the elementary excitations, i.e., with leptons, quarks and flavor-gluons (as, e.g., the `would-be' NG bosons indeed do, see Sec.~\ref{Physical_view}), the low-momentum $\bar h^{2}_{ab}(q^2)$ does not reduce to the summation of a geometric series
of the type (\ref{barh}).

In asking about the massiveness or masslessness of a theory we are
essentially asking about the infrared behavior of the propagators.
It is then natural to employ the renormalization group \cite{Stern:1976jg}.
Such an analysis implies \cite{Lane:1974us,Stern:1976jg}: (i) Theories with an infrared
stable origin do not generate masses dynamically. (ii) Theories with
a nontrivial infrared fixed point are the candidates. (iii) For dynamical
appearance of a nontrivial mass pole the canonical dimension of the
corresponding field $f$ has to be canceled by large anomalous
dimension $\gamma_f(h_{*})$ at the nontrivial infrared fixed point
$h_{*}$.

To proceed we make the low-momentum Ansatz for $\bar h^2_{ab}(q^2)$
which explicitly gives the nontrivial infrared fixed point and has certain
phenomenological appeal. Consider the identity
\begin{eqnarray}
\frac{h^2}{q^2} &=& \frac{h^2}{q^2}\Big[\big(1-\Pi(q^2)\big)^{-1}-\Pi(q^2)\big(1-\Pi(q^2)\big)^{-1}\Big]
\,. \qquad
\label{Pi}
\end{eqnarray}
Our suggestion is to use the first term in square brackets with its
corresponding $\Pi$ for high momenta (as we did), and the second
term with its corresponding $\Pi$, (\ref{PiIR}), for the low momenta.
This results in the formula for the low-momentum $\bar
h^2_{ab}(q^2)$ going at $q^2=0$ to the non-perturbative fixed point
$h_{*}^2$:
\begin{eqnarray}
\bar h^2_{ab}(q^2) &=& h_{*}^2\Big[-\Pi(q^2)\big(1-\Pi(q^2)\big)^{-1}\Big]_{ab} \,.
\label{hIR}
\end{eqnarray}
The fermion self-energies $\Sigma$ differ in different channels by
the low-momentum flavor-sensitive interaction strengths (\ref{hIR}),
basically due to the low-momentum symmetry breaking flavor-gluon
self-energy. It is noteworthy that the way to the infrared
fixed point $h_*$ is matrix-fold.

For demonstration of the fixed point we ignore the matrix character
of the problem, replace in (\ref{hIR}) $M^2_{ab}$ by $M^2$ and
compute the corresponding beta function around the non-perturbative
fixed point $h_{*}$:
\begin{eqnarray}
\beta(\bar h)\equiv 2\frac{\d\bar h}{\d \ln(-q^2/M^2)} &=&
\frac{1}{h_{*}^2}\bar h(\bar h^2-h_{*}^2) \,.
\end{eqnarray}

\section{Fermion mass generation}
\label{sec:fer_mass_gen}

The charged fermion mass $m_f$, $f=e,u,d$, or, in general, its chirality-changing
proper self-energy $\Sigma_f$ is a bridge between the right- and the
left-handed field: $\bar f_R m_f f_L$ or, in general, $\bar f_R \Sigma_f f_L$.
This is possible here, because the flavor-gluons interact both with right- and left-handed
fermion fields. Moreover, due to the representation assignments
fixed in Sec.~\ref{sec:lagrangian} the self-energies (and hence mass matrices) of fermions with different electric charges are expected to be different: $\bar e_R(\trip)\,\Sigma_e\,e_L(\atrip)\neq \bar
u_R(\trip)\,\Sigma_u\,u_L(\trip)\neq \bar d_R(\atrip)\,\Sigma_d\,d_L(\trip)$.
The matrix SD `gap' equation for $\Sigma_f$ thus takes the form  (see Fig.~\ref{fig:SDE})
\begin{eqnarray}
\Sigma_f(p^2) &=& -3\I \int\!\frac{\d^4k}{(4\pi)^4}\,
\frac{\bar h^2_{ab}(k^2)}{k^2}
\nonumber \\ && \times \,
T_{f_R,a} \, \Sigma_f(\ell^2) \Big[\ell^2-\Sigma^{\dag}_f(\ell^2)\,\Sigma_f(\ell^2)\Big]^{-1} T_{f_L,b} \,,
\nonumber \\ &&
\label{Sigmaf}
\end{eqnarray}
where $\ell \equiv p-k$. It is common in the fermion SD equations to use the transverse gauge in which the fermion wave
function renormalization can be ignored.

We expect that the solution $\Sigma_f$, if corresponds to the real world, has the following properties:
\begin{itemize}
  \item It has eigenvalues appropriate to the charged fermion mass spectrum.
  \item Its conceivable complex structure embeds the CP violating phases, and in case of quarks it reproduces the CKM matrix \cite{Benes:2009iz}.
  \item Moreover, it breaks spontaneously the axial global symmetries $\group{U}(1)_{\mathrm{B}_{5}}$ and  $\group{U}(1)_{\mathrm{L}_{5}}$, giving rise to composite NG excitations, which we discuss in Sec.~\ref{sec:experimental}.
\end{itemize}

Analogous SD equation can be written for the neutrino self-energy $\Sigma_n$, which we deliberately express in the block form
\begin{eqnarray}
\Sigma_n &=& \left(\begin{array}{cc} \Sigma_{L} & \Sigma_{D} \\ \Sigma_{D}^\T & \Sigma_{R} \end{array}\right)
\,,
\label{Sigmanu}
\end{eqnarray}
where $\Sigma_{L}$ and $\Sigma_{R}$ are symmetric matrices of dimensions $3 \times 3$ and $9 \times 9$, respectively. While the block $\Sigma_{D}$ in the decomposition \eqref{Sigmanu} corresponds to the Dirac mass term $\bar \nu_{R}^s\,\Sigma_{D}^s \, \nu_L$, analogous to those of the charged fermions, the blocks $\Sigma_{L}$ and $\Sigma_{R}$ correspond to Majorana mass terms $\bar\nu_L\,\Sigma_L\,(\nu_L)^{\C}$ and $\bar \nu_{R}^s\,\Sigma_{R}^{sr}\,(\nu_{R}^r)^{\C}$, respectively. The point is that the form of the neutrino SD equation imply, due to specific assignment of neutrinos in the $\group{SU}(3)_{\mathrm{F}}$ representations, non-vanishing $\Sigma_{L}$, $\Sigma_{R}$ and accordingly the Majorana character of the resulting neutrino mass eigenstates.

Assume that the general solution $\Sigma_n$ exists, has no accidental symmetries and is phenomenologically acceptable. The emergent picture of the neutrino sector is the following:
\begin{itemize}
  \item Upon diagonalization \cite{Bilenky:1980cx} $\Sigma_{n}=U_{n}\,m_{n}\,U_{n}^\T$, where $U_{n}$ is a unitary matrix, we obtain twelve massive Majorana neutrinos with masses given by the elements of the diagonal non-negative matrix $m_{n}$.
  \item Three left-handed neutrino fields entering the electroweak currents are their linear combinations. Consequently, the mixing matrix in the effective three-neutrino world is not unitary.
  \item The lepton number and the sterility symmetry are by assumption spontaneously broken according to
\begin{eqnarray}
U(1)_{\mathrm{B}-\mathrm{L}-\mathrm{S}}\times \group{U}(1)_{\mathrm{B}+\mathrm{L}+\mathrm{S}}\times \group{U}(1)_\mathrm{S} \times \group{SU}(3)_\mathrm{S}
\nonumber \\ \longrightarrow \group{U}(1)_\mathrm{B} \,,
\end{eqnarray}
resulting in one standard massless majoron, and $1+8$ neutrino composite scalars, so called \emph{sterile majorons}, discussed in Sec.~\ref{sec:experimental}.
\end{itemize}

\subsection{Hierarchy of mass scales}

Literally, the model has the only mass scale $\Lambda_{\mathrm{F}}$
entering the perturbative formula (\ref{hPT}). It is theoretically
arbitrary and should be fixed from one experimental datum
characterizing the onset of the strong coupling r\'{e}gime, say, of
spontaneous chiral symmetry breaking. 

Ultimately, masses of all physical excitations, both elementary and
composite, are the calculable multiples of this scale. Their description deals, however, with matrix momentum-dependent
proper self-energies of elementary gauge and fermion fields.
Therefore, there is not a direct way to determination of the gauge boson
and fermion masses from them.

Nevertheless, because of the low-momentum character of masses governed by the
non-perturbative infrared fixed point and because of non-analytic
dependence of masses upon the matrix low-momentum coupling we
think the huge amplification of scales is conceivable \cite{Miransky:1996pd}.

Just for the sake of \emph{illustrating} this assertion, we use the simplified form of the effective charge $\bar h^2(k^2)=h_{*}^2/(1-k^2/M^2)$
in (\ref{Sigmaf}) for \emph{all} momenta, and replace the Wick-rotated fermion momentum-dependent \emph{matrix} self-energies
$\Sigma(k^2)$ by a simple momentum-dependent \emph{non-matrix} Ansatz
\begin{eqnarray}\label{abAnsatz}
\Sigma(k^2) &\rightarrow& \sigma(k^2)=\frac{(a+b)m^3}{ak^2+bm^2} \,,
\end{eqnarray}
with reasonably chosen parameters $a,b\geq0$. Notice that the Ansatz satisfies $\sigma(m^2)=m$.

If we set $a=0$ then the $\sigma=m$ is a constant and the SD equation \eqref{Sigmaf} with vanishing external momentum $p^2 = 0$
turns into the algebraic equation
\begin{eqnarray}
m &=& \frac{h_{*}^2}{16\pi^2}\int_{0}^{\infty}\d k^2\frac{M^2}{k^2+M^2}
\frac{m}{k^2+m^{2}} \nonumber\\
  &=& \frac{mM^2}{M^2-m^2}\log{\frac{M^2}{m^2}} \,, \label{m}
\end{eqnarray}
which has within the approximation $\tfrac{m^2}{M^2}\ll1$ the solution of form \cite{Pagels:1979ai}
\begin{eqnarray}\label{expCritScaling}
m &=& M\exp[-8\pi^2/h_{*}^{2}] \,,
\end{eqnarray}
exhibiting the appealing exponential critical scaling, but lacking the existence of nonzero critical constant.

The nonzero critical constant occurs as a result of the self-energy UV damping encountered in the Ansatz once $a\ne0$. The analytic form of the corresponding algebraic SD equation is much more complex, therefore we write it in a not-fully explicit form and again only for vanishing external momentum. We trade the integration variable $k^2$ for dimensionless one $x$, $k^2\rightarrow M^2x$, allowing to express the SD equation with the general Ansatz \eqref{abAnsatz} in terms of dimensionless parameter $\epsilon\equiv\tfrac{m^2}{M^2}$ as
\begin{eqnarray}
m &=& m\frac{h_{*}^2}{16\pi^2}\int_{0}^{\infty}\d x
\frac{b\epsilon(ax+b\epsilon)}{(1+x)\left(x(ax+b\epsilon)+\epsilon^3(a+b)^2\right)} \nonumber\\
  &=& m\frac{h_{*}^2}{16\pi^2}\left[A(a,b)+\epsilon B(\epsilon,a,b)\right] \,,
\end{eqnarray}
where $A(a,b)$ is an $\epsilon$-independent constant and $B(\epsilon,a,b)$ is a function of $\epsilon$ regular in origin. From the integral, it is easy to see that the expression in square brackets is positive for $\epsilon>0$. Further it is a decreasing function of $\epsilon$. Near origin the equation has an approximate solution
\begin{eqnarray}\label{critScaling}
\epsilon &\approx& \frac{16\pi^2}{|B(0,a,b)|}\left(\frac{1}{h_{c}^2}-\frac{1}{h_{*}^2}\right) \,,
\end{eqnarray}
where $h_c$ is given as
\begin{eqnarray}
h_{c} &=& \sqrt{\frac{16\pi^2}{A(a,b)}}
\end{eqnarray}
and it is obviously the critical coupling constant because the solution exists only for $h>h_c$.

The critical scaling \eqref{critScaling} allows arbitrarily small fermion masses, $\epsilon\rightarrow0$, at the price of fine-tuning $h_*$ to be extremely close to $h_c$. This fine-tuning is much weaker for the exponential critical scaling \eqref{expCritScaling} which, nevertheless, lacks the critical coupling constant. We believe that the ultimate critical scaling is a combination of both behaviors illustrated above, e.g., it is of Miransky-type occurring in conformal phase transition \cite{Miransky:1996pd,Braun:2010qs}, which is exponential and exhibits at the same time the non-zero critical scaling. The two critical scalings obtained here are simply too rough to exhibit the both features simultaneously.

To demonstrate the critical amplification we assume the Miransky scaling \cite{Braun:2010qs}
\begin{eqnarray}\label{MirScaling}
m &=& M\e^{-4\pi/\sqrt{h_{*}^2-h_{c}^2}} \,.
\end{eqnarray}
With $M=10^3\, \rm TeV$ the `neutrino' mass $m_\nu = 1\,\mathrm{eV}$ is obtained for $\Delta h_{\nu}^2/4\pi=4\pi/225(\ln10)^2\doteq 0.01$, and the mass of the `top' quark $m_t=10^2\, \rm GeV$ is obtained for $\Delta h_{t}^2/4\pi=4\pi/16(\ln10)^2\doteq 0.15$. $\Delta h^2=h_{*}^2-h_{c}^2$ gives the distance of critical value for a given channel from the fixed point.


\begin{center}*\end{center}

As an example of the predictive power of the model even without the necessity of solving the SD equations we mention the relation
\begin{eqnarray}
\Sigma_e(p^2) &=& \Sigma^{\dag}_d(p^2) \,,
\label{Sigma_e=Sigma_d}
\end{eqnarray}
following directly from \eqref{Sigmaf}. It states that the mass spectra of the charged leptons and the down-type quarks should be the same. Indeed \cite{Nakamura:2010zzi}, the muon and the strange quark are almost equally heavy ($m_\mu \doteq 105.7\,\mathrm{MeV}$, $m_s \doteq 101\,\mathrm{MeV}$). However, for other particles, especially for the electron, which is roughly ten times lighter than the down quark ($m_e \doteq 0.51\,\mathrm{MeV}$, $m_d \doteq 4.1-5.8\,\mathrm{MeV}$), the relation \eqref{Sigma_e=Sigma_d} does not work too well. This is however understandable, since the SD equation \eqref{Sigmaf} itself, and consequently also the relation \eqref{Sigma_e=Sigma_d}, as neglecting, e.g., the fermion wave function renormalizations and the color charge of the quarks, is only approximate.

Due to the special character of neutrinos (higher number of their right-handed components and Majorana components of their self-energy) there is no similar relation connecting $\Sigma_n$ and $\Sigma_u$.

\section{Electroweak symmetry breaking}
\label{sec:EW_sym_bre}

Once the $\group{SU}(3)_\mathrm{F}$ symmetry is spontaneously broken the fermion
self-energies inevitably break also the electroweak symmetry
$\group{SU}(2)_{\mathrm{L}} \times \group{U}(1)_{\mathrm{Y}}$ down to the electromagnetic
$\group{U}(1)_{\mathrm{em}}$. One expects the gauge bosons $W^\pm$, $Z$ to obtain
masses proportional to the fermion self-energies.

This section is devoted to mere presentation of the resulting masses without detailed derivation, as the basic reasoning is rather standard \cite{Freundlich:1970kn,Jackiw:1973tr,Cornwall:1973ts}: The gauge boson mass matrix is obtained as the residue of the massless pole of the corresponding polarization tensor, which is calculated at one-loop level. While one of the two vertices in the loop is bare, the other must satisfy the WT identity, consistent with the fermion symmetry-breaking propagators, so that the polarization tensor is transversal. As a `side-effect' of imposing the WT identity the vertex will contain a massless pole, proportional to the symmetry-breaking fermion self-energies. This pole is to be interpreted as the propagator of the `would-be' NG boson, giving mass to the corresponding gauge boson.

The resulting gauge boson masses squared can be written in form of the sum rules as
\begin{subequations}
\label{eq:EW:sum_rules}
\begin{eqnarray}
M_Z^2 &=& (g^2+g^{\prime2})(\mu_{u}^2+\mu_{d}^2+\mu_{\nu}^2+\mu_{e}^2) \,,
\\
M_W^2 &=& g^2(\mu_{q}^2+\mu_{\ell}^2) \,,
\end{eqnarray}
\end{subequations}
with the individual contributions given by (the arguments $p^2$ at the $\Sigma$s are suppressed)
\begin{widetext}
\begin{subequations}
\label{eq:EW:sum_rules:mu}
\begin{eqnarray}
\mu_{f=u,d,e}^2 &=&  -\I\frac{1}{2}  \int \frac{\d^4 p}{(2\pi)^4}
\Tr\bigg\{ \Sigma_f^\dag\Sigma_f^{\vphantom{\dag}} \, a_f^2 -
\frac{1}{2} p^2
\big(\Sigma_f^\dag\Sigma_f^{\vphantom{\dag}}\big)^\prime a_f^2
\bigg\} \,,
\label{eq:EW:sum_rules:mu_ude}
\\
\mu_{\nu}^2 &=& -\I\frac{1}{2} \int\!\frac{\d^4
p}{(2\pi)^4}\Tr\bigg\{ \Big(
PP^\T\,\Sigma_n^\dag\,\Sigma_n^{\vphantom{\dag}} +
\Sigma_n^\dag\,PP^\T\,\Sigma_n^{\vphantom{\dag}} \Big)\,
a_n\,PP^\T\,a_n
-\frac{1}{2}p^2\Big(\Sigma_n^\dag\Sigma_n^{\vphantom{\dag}}\,PP^\T+
\Sigma_n^\dag\,PP^\T\,\Sigma_n^{\vphantom{\dag}}\Big)^\prime\,a_n\,PP^\T\,a_n
\nonumber \\ &&
\hphantom{-\I\frac{1}{2} \int\!\frac{\d^d p}{(2\pi)^d}\Tr\bigg\{}
+\frac{1}{2}p^2 \Big(
PP^\T\,\Sigma_n^\dag\,\Sigma_n^{\vphantom{\dag}} -
\Sigma_n^\dag\,PP^\T\,\Sigma_n^{\vphantom{\dag}} \Big)\,
a_n\,PP^\T\,a_n^\prime \bigg\} \,,
\label{eq:EW:sum_rules:mu_nu}
\\
\mu_{q}^2 &=& -\I\frac{1}{2} \int\!\frac{\d^4 p}{(2\pi)^4}
\Tr\bigg\{ \,\,\phantom{+\frac{1}{2} p^2}
\Big[a_u \, \Sigma_u^\dag
\Sigma_u^{\vphantom{\dag}} \, a_d + a_d \, \Sigma_d^\dag
\Sigma_d^{\vphantom{\dag}} \, a_u\Big] -\frac{1}{2} p^2 \Big[a_u
\big(\Sigma_u^\dag \Sigma_u^{\vphantom{\dag}}\big)^\prime a_d + a_d
\big(\Sigma_d^\dag \Sigma_d^{\vphantom{\dag}}\big)^\prime a_u\Big]
\nonumber \\ && \hphantom{-\I\frac{1}{2} \int\!\frac{\d^4 p}{(2\pi)^4}\Tr\bigg\{}
+\frac{1}{2} p^2 \Big[a_u \, \Sigma_u^\dag
\Sigma_u^{\vphantom{\dag}} \, a_d^\prime + a_d \, \Sigma_d^\dag
\Sigma_d^{\vphantom{\dag}} \, a_u^\prime\Big] -\frac{1}{2} p^2
\Big[a_u^\prime \, \Sigma_u^\dag \Sigma_u^{\vphantom{\dag}} \, a_d +
a_d^\prime \, \Sigma_d^\dag \Sigma_d^{\vphantom{\dag}} \, a_u\Big]
\bigg\} \,,
\label{eq:EW:sum_rules:mu_q}
\\
\mu_{\ell}^2 &=& -\I\frac{1}{2} \int\!\frac{\d^4p}{(2\pi)^4}
\Tr\bigg\{ \,\,\phantom{-\frac{1}{2} p^2\,}
\Big[a_n \, \Sigma_n^{\dag} \Sigma_n^{\vphantom{\dag}} \, P \, a_e
\, P^\T + a_e \, \Sigma_e^{\dag} \Sigma_e^{\vphantom{\dag}} \,
P^\T \, a_n \, P \Big]
\nonumber \\ &&
\hphantom{-\I\frac{1}{2} \int\!\frac{\d^4 p}{(2\pi)^4}\Tr\bigg\{}
-\frac{1}{2} p^2 \Big[a_n \, \big(\Sigma_n^{\dag}
\Sigma_n^{\vphantom{\dag}}\big)^\prime P \, a_e \,
P^\T + a_e \, \big(\Sigma_e^{\dag}
\Sigma_e^{\vphantom{\dag}}\big)^\prime P^\T \, a_n \,
P \Big] \nonumber \\ && \hphantom{-\I\frac{1}{2}
\int\!\frac{\d^4 p}{(2\pi)^4}\Tr\bigg\{} +\frac{1}{2} p^2 \Big[a_n
\, \Sigma_n^{\dag} \Sigma_n^{\vphantom{\dag}} \, P \,
a_e^\prime \, P^\T + a_e \, \Sigma_e^{\dag}
\Sigma_e^{\vphantom{\dag}} \, P^\T \, a_n^\prime \,
P \Big] \nonumber \\ && \hphantom{-\I\frac{1}{2}
\int\!\frac{\d^4 p}{(2\pi)^4}\Tr\bigg\{} -\frac{1}{2} p^2
\Big[a_n^\prime \, \Sigma_n^{\dag} \Sigma_n^{\vphantom{\dag}}
\, P \, a_e \, P^\T + a_e^\prime \, \Sigma_e^{\dag}
\Sigma_e^{\vphantom{\dag}} \, P^\T \, a_n \, P \Big]
\bigg\} \,,
\label{eq:EW:sum_rules:mu_ell}
\end{eqnarray}
\end{subequations}
\end{widetext}
where the prime denotes the differentiation with respect to
$p^2$ and where we denoted ($f=u,d,n,e$)
\begin{eqnarray}
a_f &\equiv&
\big(p^2-\Sigma_f^\dag\Sigma_f^{\vphantom{\dag}}\big)^{-1}
\end{eqnarray}
and
\begin{eqnarray}
P &\equiv& \left(\begin{array}{l} 1_{3 \times 3} \\ 0_{9 \times 9} \end{array}\right) \,.
\end{eqnarray}
The r\^{o}le of the projector $P$ is to ensure that the
right-handed neutrinos contribute to the $W^\pm$, $Z$ masses only
indirectly, through the mixing with the left-handed ones, i.e.,
through the `denominator' $a_n$ of the full neutrino propagator.

Notice that while the $W^\pm$, $Z$ masses are calculated \emph{non-perturbatively} within the
g.f.d.~(through the non-perturbative fermion self-energies),
they are also at the same time calculated \emph{perturbatively} within
the electroweak dynamics (they actually are of the lowest possible, i.e., second
order in $g$, $g^\prime$).

The formul{\ae} similar to those \eqref{eq:EW:sum_rules:mu_ude}, \eqref{eq:EW:sum_rules:mu_q} for $\mu^2_{f=u,e,d}$, $\mu^2_{q}$ have been already presented in the literature \cite{Miransky:1988xi} as a straightforward generalization of the Pagels--Stokar result \cite{Pagels:1979hd}. The present formul{\ae}, however, differs from those in \cite{Miransky:1988xi} by the factor at the terms proportional to $(\Sigma^\dag\Sigma)^\prime$: While here we have the factor of $1/2$, in Ref.~\cite{Miransky:1988xi} there is rather $1/4$. As this problem is more general and does not apply only to the present particular model of g.f.d., a separate paper is to be published on this issue.

The relation $\rho \equiv M_W^2/M_Z^2\cos^2\theta_W=1$ is not guaranteed automatically. It would be certainly satisfied (as can be seen by inspection of the sum rules \eqref{eq:EW:sum_rules} together with the terms \eqref{eq:EW:sum_rules:mu}) in the case of the exact custodial symmetry, i.e., when $\Sigma_u=\Sigma_d$, $\Sigma_D=\Sigma_e$ and $\Sigma_L=\Sigma_R=0$ (provided, of course, that the numbers of the right-handed and left-handed neutrinos would be the same). This is obviously not true in the present model. Still, however, the relation $\rho=1$ can be fulfilled at least approximately, as only Higgs \emph{doublets} enter the effective Lagrangian introduced in Sec.~\ref{Physical_view} \cite{Chanowitz:1985ug}. We obtain this way an additional possibility how to, in principle, test the model.

\section{Experimental consequences}
\label{sec:experimental}

In this section we list the generic phenomena which the model
predicts regardless of details of the spectrum of its elementary
excitations (fermions and gauge bosons) which we are not able to
reliably compute anyway. Detailed discussion of the relevance of
these phenomena to reality is postponed to future work.


\subsection{Scalars from spontaneously broken global symmetries}

The model has six global symmetry currents defined in Sec.~\ref{sec:lagrangian}. At
quantum level two global symmetries $\group{U}(1)_{\mathrm{B}-(\mathrm{L}+\mathrm{S})}$ and $\group{SU}(3)_\mathrm{S}$
are exact. The remaining four global Abelian symmetries are
explicitly broken by four distinct (electroweak, QCD and g.f.d.)
anomalies.

Assume that the most general chirality-changing fermion proper
self-energies of all fermion fields,
\begin{equation}
\Sigma_u \,,\quad \Sigma_d \,,\quad
\Sigma_n=\beginm{cc} \Sigma_L & \Sigma_D \\ \Sigma_{D}^\T & \Sigma_R \endm \,,\quad
\Sigma_e \,,
\end{equation}
introduced in Sec.~\ref{ssec:fermions}, are dynamically generated. I.e., no additional
selection rule is at work. As a result different symmetries generated
by charges of the corresponding currents are spontaneously broken by
different self-energies differently:
\begin{equation}
\begin{array}{lll}
\ \ \ B & \mbox{is not broken,} &  \\
\ \ \ B_{5} & \mbox{is broken by} & \Sigma_u\,,\ \Sigma_d\,, \\
\ \ \ L & \mbox{is broken by} & \Sigma_L\,,\ \Sigma_D\,, \\
\ \ \ L_{5} & \mbox{is broken by} & \Sigma_e\,,\ \Sigma_L\,,\ \Sigma_D\,, \\
\ \ \ S & \mbox{is broken by} & \Sigma_R\,,\ \Sigma_D\,, \\
SU(3)_{\mathrm{S}} & \mbox{is broken by} & \Sigma_R\,,\ \Sigma_D\,.
\end{array}
\end{equation}
It is then irrefutable that five types of the (pseudo-) NG collective excitations
emerge in the spectrum of the model.

Due to the anomalies it is not easy to link the scalars with given
spontaneously broken currents. Therefore it is not easy to say what
physical characteristics, like mass, the scalars will have. The
structure of anomalies translates into a mass matrix of the scalars,
and only after the diagonalization of the mass matrix, the mass
spectrum of the scalars and their couplings to currents are revealed.
We believe that the mass spectrum reflects the hierarchy
of scales that are given by dynamics of individual anomalies.

In the following we merely list the scalars. More comprehensive
discussion of this important issue will be presented in separate paper.

\subsubsection{Massless Abelian Majoron $J$}

The exact, anomaly free Abelian symmetry $\group{U}(1)_{\mathrm{B}-(\mathrm{L}+\mathrm{S})}$ is
spontaneously broken by $\Sigma_n$. It gives rise to the
\emph{massless composite Abelian Majoron} $J$ \cite{Chikashige:1980ui,Schechter:1981cv},
\begin{eqnarray}
m_J &=& 0 \,.
\end{eqnarray}
Its coupling
to the observed fermions is heavily suppressed by the scale where the
neutrino self-energy $\Sigma_n$ is generated and the corresponding
symmetry is broken. At the present exploratory level, we identify the
scale with $\Lambda_{\mathrm{F}}$.

\subsubsection{Octet of massless Majorons $J_\sigma$}

By assumption the exact global non-Abelian $\group{SU}(3)_\mathrm{S}$ sterility symmetry is
completely spontaneously broken by $\Sigma_D$, and $\Sigma_R$. It gives rise
to an \emph{octet of massless composite Majorons} $J_\sigma$, $\sigma=1,\ldots,8$:
\begin{eqnarray}
m_{J_\sigma} &=& 0 \,.
\end{eqnarray}
Their coupling to the observed fermions is even more suppressed because they are predominantly an admixture of the sterile neutrinos. Moreover, we stress the fact \cite{Gelmini:1982zz} that massless Majorons do not imply a new long-range force as the induced potential between two fermions is spin-dependent and tensorial, with a $1/r^3$ fall off \cite{Peccei:1998jv}.

%

\subsubsection{Light leptonic axion $a^\prime$}

The current $L_{5}+3L$ is a typical linear combination of symmetry
currents whose divergence is dominated by electroweak anomalies.
Spontaneous breakdown of such a `would-be', still rather good,
symmetry by lepton masses implies an almost massless \emph{superlight}
composite pseudo-NG boson, known in the literature as the
\emph{leptonic axion or arion $a^\prime$} \cite{Anselm:1981aw}. Its tiny mass
estimate is \cite{Anselm:1990uy}
\begin{eqnarray}
m_{a^\prime} &\approx& \e^{-\frac{4\pi^2}{g^2}}M_W\approx 10^{-42}M_W \,,
\end{eqnarray}
where we used value \cite{Nakamura:2010zzi} of the weak coupling constant $g \doteq 0.653$.

\subsubsection{Light Weinberg-Wilczek axion $a$}

The currents $B_{5}-4S$ or $B_{5}-(L_{5}-L)$ are the typical linear
combinations of symmetry currents whose divergences are dominated
by the QCD anomaly. Consequently, their corresponding charges are
the natural candidates for the generators of the Peccei--Quinn symmetry
\cite{Peccei:1977hh,*Peccei:1977ur}. Spontaneous breakdown of this `would-be' symmetry
by the fermion (both quark and lepton) proper self-energies implies
the \emph{Weinberg--Wilczek axion} $a$ \cite{Weinberg:1977ma,Wilczek:1977pj}. Due to
its compositeness caused by the flavor dynamics the axion is invisible
with mass \cite{Kim:1984pt}
\begin{eqnarray}
m_{a} &\approx& \frac{m_\pi f_\pi}{\Lambda_{\mathrm{F}}} \,.
\label{mass_a}
\end{eqnarray}
At this point we in fact \emph{fix} the scale $\Lambda_\mathrm{F}$: The mass estimate \eqref{mass_a} together with the invisibility of the axion imply stringent restriction $\Lambda_\mathrm{F} \sim 10^9-10^{12}\,\mathrm{GeV}$ \cite{Raffelt:2006rj}. Such a high scale guarantees that the axion is light enough, and it interacts with fermions weakly enough so that it could not have been observed so far.

The QCD anomaly induces direct vertex of the axion $a$ with gluons
\begin{eqnarray}
{\cal L}_{agg} &\propto&
\frac{g_\mathrm{s}^2}{32\pi^2}\frac{a}{\Lambda_{\mathrm{F}}} G \tilde G \,,
\end{eqnarray}
necessary to eliminate the QCD $\theta$-term via the Peccei--Quinn
mechanism \cite{Peccei:1977hh,*Peccei:1977ur}.

\subsubsection{Superheavy Majoron $H$}

Consider a linear combination of the currents $L$, $L_{5}$ and $S$ whose
divergences are dominated by the strong g.f.d.~anomaly. Spontaneous
breakdown of such a `would-be' symmetry by lepton masses implies
yet another collective excitation, the \emph{superheavy sterile
majoron} $H$.

The majoron $H$ acquires huge mass due to the strong flavor axial
anomaly. Its mass can be estimated
according to the $\eta^\prime$ mass analysis in QCD \cite{Witten:1979vv,Veneziano:1979ec} as
\begin{eqnarray}
m_{H} &\approx& \Lambda_{\mathrm{F}} \,.
\end{eqnarray}

The anomalous coupling of $H$ to the flavor gauge bosons is given as
\begin{eqnarray}\label{axionEWinteraction}
{\cal L}_{HCC} &\propto&
\frac{h^2}{32\pi^2}\frac{H}{\Lambda_{\mathrm{F}}} F \tilde F \,.
\end{eqnarray}
Due to this interaction the g.f.d.~$\theta$-term is eliminated via the
Peccei--Quinn mechanism \cite{Peccei:1977hh,*Peccei:1977ur}.

The superheavy sterile majoron $H$, composed predominantly of the right-handed neutrinos, evokes the cosmological inflation scenario driven by right-handed neutrino condensate \cite{Barenboim:2008ds,Barenboim:2010nm}.

\subsection {Sterile right-handed neutrino fields}

Sterile neutrinos were introduced for purely theoretical reason of
anomaly freedom. Predicting the neutrino spectrum is an unsolved,
hard, challenging dynamical problem. Here we only mention that the
sterile neutrinos with masses in the $\mathrm{keV}$ range become
increasingly phenomenologically welcome as the candidates for dark
matter \cite{Nieuwenhuizen:2008pf,Kusenko:2009up,Bezrukov:2009th}.

\section{Comparison with other models}
\label{sec:comparison}

Before we conclude we want to demarcate the g.f.d.~model against other similar models (especially the ETC models). It is worth stating here clearly which of its ingredients have been already used elsewhere and which are new. Let us start with the former ones:
\begin{itemize}
\item The very idea of gauging the flavor is of course not new \cite{Wilczek:1978xi,Ong:1978tq,Davidson:1979wt,Chakrabarti:1979vy,Yanagida:1979gs,Yanagida:1980xy,Nagoshi:1990wk,Berezhiani:1990wn} and is inherent also in the ETC models. Common feature of all such models, embedding the observed fermions into the complex flavor representations, is the necessity to compensate their gauge anomaly by postulating the existence of new fields, most simply electroweak singlet fermions, the right-handed neutrinos.

\item The specific $\group{SU}(3)_{\mathrm{F}}$ representation assignment of chiral fermions, employed within the g.f.d.~model, was used already in certain ETC models \cite{Appelquist:2003hn}, with the same aim to distinguish among the magnitudes of masses of fermion with different electric charges.

\item The self-breaking gauge dynamics was studied by Eichten and Feinberg \cite{Eichten:1974et}, pursued further by Pagels \cite{Pagels:1979ai} and used in tumbling models \cite{Raby:1979my,Martin:1992aq} and in some ETC models \cite{Ryttov:2010kc}.

\item In the g.f.d.~model, in contrast to ETC models, there is no residual confining gauge dynamics responsible for EWSB by means of generating technifermion condensates in the QCD manner. It is directly the standard fermion masses, top-quark mass in particular, which plays the r\^{o}le of technifermion EWSB condensate. The g.f.d.~model resembles in this respect the top-quark condensate models \cite{Miransky:1988xi}.

\item Already Pagels \cite{Pagels:1979ai} attempted to demonstrate that the arbitrarily small fermion masses can be achieved within the models of chiral gauge symmetry breaking. The infrared effective coupling constant of g.f.d.~has to be found above but close enough to its critical value. The amplification of scales within, e.g., walking ETC models is caused by significant slow-down of the effective coupling constant evolution due to the proximity of the `would-be' infrared fixed point after it surpasses its critical value \cite{Holdom:1981rm}.
\end{itemize}

The g.f.d.~model brings the fusion of these appealing quantum field theoretical features by encompassing them naturally within the rigid framework and arranging them in a distinctive and individual way:
\begin{itemize}
\item First of all, the g.f.d.~model is economical in the sense that it introduces only the octet of flavor-gluons and a limited number of right-handed neutrinos. The only new tunable free parameter is the $\group{SU}(3)_{\mathrm{F}}$ coupling constant.

\item The g.f.d.~model can be viewed as a UV complete dynamics that makes the top-quark and eventually other fermions to condense. In contrast to the top-quark condensate models, however, the g.f.d.~mass generation is not governed by an UV fixed point (which would not be, after all, in the spirit of Stern's analysis \cite{Stern:1976jg}), but it is rather associated with the existence of an IR fixed point, not in contradiction with the conclusion of Stern. In both types of models the canonical dimensions of fields are canceled by large anomalous dimensions at the nontrivial fixed point.

\item The arbitrary smallness of fermion masses has its origin in the critical behavior in the vicinity of the fixed point. The evolution of the effective coupling constant not only slows down, but saturates by the true infrared fixed point just above the critical value.

\item Due to the fact that the flavor symmetry is broken, the low-momentum running of the effective flavor coupling towards its non-perturbative IR fixed point is effectively matrix-fold. The wild fermion mass spectrum is an imprint of this matrix structure, amplified by the critical behavior near the critical point.
\end{itemize}

\section{Conclusions}
\label{sec:conclusions}

Present model of soft generation of masses of the SM particles by
a strong-coupling g.f.d.~is rather rigid and, if reliably solved,
easily falsifiable. (i) It has just one unknown parameter
$h$ which by dimensional transmutation converts into the
(theoretically arbitrary) mass scale $\Lambda_{\mathrm{F}}$. This implies
that the dynamically generated masses are related. Ultimately, this
fact provides experimental tests and predictions of the
model even without resorting to high energies. (ii) One elaborated
example of mass relations is the sum rules for the electroweak
boson masses $M_Z$, $M_W$, Eq.~\eqref{eq:EW:sum_rules}. The implication
of these sum rules is interesting: \emph{There is no generic electroweak
scale in the model}. The electroweak boson masses are merely a
manifestation of the large top quark mass \cite{Hosek:1982cz,Hosek:1985jr,Miransky:1988xi,Bardeen:1989ds}.
(iii) The neutrino right-handed singlets are introduced not in order to
describe the experimental fact of massiveness of the neutrinos. They
are enforced by the purely theoretical requirement of the absence of
axial anomalies. Explicit computation of the neutrino mass spectrum
represents an alluring prediction and a crystalline theory challenge.
The very existence of sterile neutrinos introduced for anomaly freedom
should have experimental consequences in neutrino oscillations and in
astrophysics. (iv) The fermion SD equations fix also
the fermion mixing parameters. (v) It is natural to expect that the
unitarization of the scattering amplitudes of the longitudinal
polarization states of massive spin one particles proceeds in the
present model via the massive composite `cousins' of the composite
`would-be' NG bosons. Its practical implementation is obscured,
however, by our ignorance of the detailed properties of the spectrum
of strongly coupled $\group{SU}(3)_\mathrm{F}$.

We believe that the physical arguments which yield the dynamical mass
generation of the SM particles are robust and respect the quantum field
theory common sense. It is perhaps noteworthy that the model provides,
regardless of our will, natural candidates also for dark matter: sterile
neutrinos, axions and majorons. We cannot exclude, however, that the world
which the model describes is, numerically, not ours.

\begin{acknowledgments}
We are grateful to Ji\v{r}\'{i} Ho\v{r}ej\v{s}\'{i}, Maxim Y.~Khlopov and Ji\v{r}\'{i} Novotn\'{y} for fruitful discussions. The work was supported by the Grant LA08015 of the Ministry of Education of the Czech Republic.
\end{acknowledgments}

\appendix

\section{Fermion field content}
\label{appendix}

On the one hand, the anomaly freedom allows only specially balanced right-handed neutrino settings. On the other hand, we may not add too many right-handed neutrinos in order not to spoil the asymptotic freedom of the flavor dynamics.

Whether the model is asymptotically free is given by the negativity of the one-loop $\beta$-function
\begin{eqnarray}
\label{beta}
\beta(h)
&=& -\frac{h^3}{(4\pi)^2}\left[\frac{11}{3}C(8)-\frac{2}{3}N^{\mathrm{ew}}C(3)
-\frac{2}{3}\eta_{\mathrm{AF}}\right] \nonumber \\
&=& -\frac{h^3}{(4\pi)^2}\left[6 - \frac{2}{3}\eta_{\mathrm{AF}}\right] \,.
\end{eqnarray}
where
\begin{eqnarray}
\label{etaAF}
\eta_{\mathrm{AF}} &\equiv& \sum_r N^{\nu_R}_r\,C(r) \nonumber \\
&=& \tfrac{1}{2}N^{\nu_R}_3+\tfrac{5}{2}N^{\nu_R}_6+3N^{\nu_R}_8+\tfrac{15}{2}N^{\nu_R}_{10}+\ldots \qquad
\end{eqnarray}

The coefficient $C(r)$ reflects \emph{the flavor symmetry representation}
of the corresponding field, and is related to the quadratic Casimir
invariant $C_2(r)$. Their definitions and their relations are
\begin{eqnarray}
\delta^{ab} \, C(r) & = & \Tr{T^{a}_r \, T^{b}_r} \,, \label{coef}\\
d(r)\,C_2(r) & = & \Tr{T^{a}_r \, T^{a}_r} \,, \label{quadraticCasimir}\\
d(r)\,C_2(r)  & = & d(G)\,C(r) \,.
\end{eqnarray}
The values of the coefficients for few lowest representations are
listed in Tab.~\ref{table}.
\begin{table}[t]
\begin{tabular}{lc|cc|cc}
$r$ & $d(r)$ & $C(r)$ & $C_2(r)$ & $A(r)$ & $C_3(r)$ \\
\hline
\hline
$\mathbf{3}(\overline{\mathbf{3}})$   & $3$  & $1/2$  & $4/3$  & $(-)1$  & $(-)10/9$ \\
$\mathbf{6}(\overline{\mathbf{6}})$   & $6$  & $5/2$  & $10/3$ & $(-)7$  & $(-)35/9$ \\
$\mathbf{8}$                          & $8$  & $3$    & $3$    & $0$     & $0$       \\
$\mathbf{10}(\overline{\mathbf{10}})$ & $10$ & $15/2$ & $6$    & $(-)27$ & $(-)9$ \\
\hline
\hline
\end{tabular}
\caption{\small We list important coefficients for the lowest representations of the group $\group{SU}(3)$. The coefficient $C(r)$ and the quadratic Casimir invariant $C_2(r)$ are defined in \eqref{coef} and \eqref{quadraticCasimir}, respectively. The anomaly coefficient $A(r)$ and the cubic Casimir invariant $C_3(r)$ are defined in \eqref{CasimirI} and \eqref{anomalyC}, respectively.}
\label{table}
\end{table}

The consistence of the g.f.d.~model requires that the divergence of the flavor current
\begin{eqnarray}
\partial_\mu j^{\mu}_a= -\frac{h^2}{64\pi^2}d_{abc}F^{\mu\nu}_b\tilde{F}_{c\mu\nu}\sum_f A(r_f)
\end{eqnarray}
vanishes.

The anomaly coefficients $A(r)$ are given by
\begin{eqnarray}
\frac{1}{2}d^{abc}A(r) & = & \Tr{T^{a}_r\{T^{b}_r,T^{c}_r\}} \,, \label{anomalyC}\\
d(r) \, C_3(r) & = & d^{abc}\Tr{T^{a}_r\,T^{b}_r\,T^{c}_r} \,, \label{CasimirI}\\
2 d(r) \, C_3(r)  & = & \frac{5}{6}\,d(G)\,A(r) \,.
\end{eqnarray}
The values for some of the lowest representations are listed in Tab.~\ref{table}.

The anomaly contributions of all electroweakly charged fermions of the two cases I and II of the
model are listed in Tab.~\ref{f_setting}.

For completeness, recall that an overall complex conjugate of the representations
in the two settings is also acceptable, but it is merely a matter of
convention which multiplets we denote as a triplet and which as an
antitriplet.

From Tab.~\ref{f_setting} we see that we need to compensate 3 (5) units of triplet anomaly
coefficient. The simplest solution is to add 3 (5) triplets of
right-handed neutrinos. But instead of this minimal setting we
can add also some balanced, more complicated set including higher
representations. Constructing such non-minimal versions of the
model notice that a pair of $r$, and $\overline{r}$, as well as
real representations do not contribute to the anomaly.

The number of right-handed neutrinos is constrained by the inequality
\begin{eqnarray}
\eta_{\mathrm{AF}} &<& 9 \,,
\label{etaAF_inequality}
\end{eqnarray}
(see definition \eqref{etaAF} of $\eta_{\mathrm{AF}}$) which follows directly from the requirement of negativity of the one-loop $\beta$-function \eqref{beta}.

If the right-handed neutrino setting contained a decuplet or higher
complex representation, we would need to compensate its high anomaly
coefficient $A(r)\geq 27$ by too many other multiplets so that
the $\beta$-function would become positive. Adding real representations higher than
octet immediately makes the $\beta$-function positive too. So we are allowed to combine
only lower representations, $\mathbf{3}$, $\overline{\mathbf{3}}$, $\mathbf{6}$,
$\overline{\mathbf{6}}$, and $\mathbf{8}$ and \emph{nothing else}. Tab.~\ref{AAfreeSettings} shows all possible asymptotically and anomaly free settings.

\begin{table*}[t]
\begin{tabular}{c|l|l|l}
case &  \ settings of $\nu_R$ $\group{SU}(3)_\mathrm{F}$ representations ($n,m,k\geq0$) & \qquad $\eta_{\mathrm{AF}}$ & \qquad $\#$ of $\nu_R$ \\
\hline
\hline
I & $3\times\mathbf{3} \phantom{{}+12\times\overline{\mathbf{3}}}\quad + n\times(\mathbf{3}+\overline{\mathbf{3}})+m\times\mathbf{8}+k\times(\mathbf{6}+\overline{\mathbf{6}})$
& $3/2\phantom{1}\ +(n+3m+5k\leq7)$  & $9\phantom{5}\ +6n+8m+12k$ \\
II & $5\times\mathbf{3} \phantom{{}+12\times\overline{\mathbf{3}}} \quad+n\times(\mathbf{3}+\overline{\mathbf{3}})+m\times\mathbf{8}+k\times(\mathbf{6}+\overline{\mathbf{6}})$
& $5/2\phantom{1}\ +(n+3m+5k\leq6)$ & $15\ +6n+8m+12k$ \\
II & $1\times\mathbf{6}+2\times\overline{\mathbf{3}}\phantom{1} \quad+n\times(\mathbf{3}+\overline{\mathbf{3}}) +m\times\mathbf{8}+k\times(\mathbf{6}+\overline{\mathbf{6}})$
& $7/2\phantom{1}\ +(n+3m+5k\leq5)$ & $12\ +6n+8m+12k$ \\
I & $1\times\mathbf{6}+4\times\overline{\mathbf{3}}\phantom{1}\quad+n\times(\mathbf{3}+\overline{\mathbf{3}})+m\times\mathbf{8}$
& $9/2\phantom{1}\ +(n+3m\leq4)$ & $18\ +6n+8m$ \\
I & $1\times\overline{\mathbf{6}}+10\times\mathbf{3}\quad+n\times(\mathbf{3}+\overline{\mathbf{3}})$
& $15/2\ +(n\leq1)$          & $36\ +6n$ \\
II & $1\times\overline{\mathbf{6}}+12\times\mathbf{3}$ & $17/2$ & $42$ \\
\hline
\hline
\end{tabular}
\caption{\small The list of all possible anomaly and asymptotically free settings of $\group{SU}(3)_\mathrm{F}$ right-handed neutrino representations. Since the anomaly freedom is not affected by adding any number of pairs of conjugate representations or any number of real representations, the possibility of adding $n$ pairs of $\trip$ and $\atrip$, $m$ copies of $\mathbf{8}$ and $k$ pairs of $\mathbf{6}$ and $\overline{\mathbf{6}}$ is indicated in the second column. However, adding too many of such representations would spoil the asymptotic freedom, hence the third column shows for each setting the corresponding value of $\eta_\mathrm{AF}$, \eqref{etaAF}. The indicated inequalities correspond to the asymptotic freedom condition $\eta_\mathrm{AF}<9$, \eqref{etaAF_inequality}. Clearly, as the numbers $n$, $m$, $k$ are integers, there is consequently only a finite number of all possible settings.}
\label{AAfreeSettings}
\end{table*}

\bibliography{references}

\end{document}